\documentclass[pre,tighten]{revtex4}

\usepackage{amssymb,amsmath}

\usepackage{graphicx}

\begin{document}
\title{On the Einstein relation in a heated granular gas}
\author{Vicente Garz\'{o}}
\email[E-mail: ]{vicenteg@unex.es}
\address{Departamento de F\'{\i}sica, Universidad de Extremadura, E-06071
Badajoz, Spain}


\begin{abstract}

Recent computer simulation results [Barrat {\em et al.}, Physica A 334 (2004) 513] for granular mixtures subject to stochastic driving have shown the validity of the Einstein relation $\epsilon\equiv D/(T_0\lambda)=1$ between the diffusion $D$ and mobility $\lambda$ coefficients when the temperature of the gas $T$ is replaced by the temperature of the impurity $T_0$ in the usual Einstein relation. This problem is analyzed in this paper by solving analytically the Boltzmann-Lorentz  equation from the Chapman-Enskog  method. The gas is heated by the action of an external driving force (thermostat) which does work to compensate for the collisional loss of energy. Two types of thermostats are considered: (a) a deterministic force proportional to the particle velocity (Gaussian thermostat), and (b) a white noise external force (stochastic thermostat). The diffusion and mobility coefficients are given in terms of the solutions of two linear integral equations, which are approximately solved up to the second order in a Sonine polynomial expansion. The results show that the violation of the Einstein relation ($\epsilon\neq 1$) is only due to the non-Maxwellian behavior of the impurity velocity distribution function (absence of the Gibbs state). At a quantitative level, the kinetic theory results also show that the deviation of $\epsilon$ from 1 is more significant in the case of the Gaussian thermostat than in the case of the stochastic one, in which case the deviation of the Einstein relation is in general smaller than 1\%. This conclusion agrees quite well with the  results found in computer simulations.

Keywords: Granular gases; Thermostat forces; Kinetic Theory; Mobility and diffusion coefficients.

\end{abstract}

\draft
\pacs{ 05.20.Dd, 45.70.Mg, 51.10.+y, 47.50.+d}

\date{\today}
\maketitle


\section{Introduction}
\label{sec1}

In spite of the growing theoretical support for the validity of the hydrodynamic description to granular fluids, some care is warranted in translating properties of normal fluids to those with inelastic collisions. Thus, for instance, in the case of an {\em impurity} particle immersed in an {\em unforced}  granular gas the response to an external force on the impurity particle leads to a mobility coefficient that violates the Einstein relation \cite{McL89} between the mobility and diffusion coefficients. This violation has been recently studied by Dufty and Garz\'o \cite{DG01} by using linear response methods and without any limitation on the mass ratio, density, or degree of inelasticity. The analysis shows that there are three different reasons why the Einstein relation does not apply in granular gases: the absence of the Gibbs state, the cooling of the reference state, and the occurrence of different kinetic temperatures between the impurity and gas particles.   

However, very recently Barrat {\em et al.} \cite{BLP04} have performed computer simulations to test  the validity of fluctuation-dissipation relations for a mixture constituted by two granular gases with different masses and restitution coefficients. The system is driven by an external energy input to achieve a stationary state. Specifically, they use a homogeneous stochastic driving ({\em thermostat}) in the form of a white noise plus a friction term. Given that the partial granular temperatures of each species are different, they observe that the classical equilibrium Green-Kubo relations are satisfied separately for each component, the role of the ``equilibrium'' temperature being played by the granular temperature of each component. In particular, in the limit in which the mole fraction of one of the species is negligible (tracer limit),  the Einstein relation holds when the temperature of the gas $T$ is replaced by the temperature of the impurity $T_0$. In this case, the (modified) Einstein ratio between the diffusion coefficient $D$ and the mobility coefficient $\lambda$ is verified, namely, 
\begin{equation}
\label{n1}
\epsilon\equiv \frac{D}{T_0\lambda}=1.
\end{equation}

The results obtained by Barrat {\em et al.} \cite{BLP04} from computer simulations motivate this work.  The aim of this paper is to determine the Einstein ratio $\epsilon$ for a {\em driven} granular gas in the context of the Boltzmann equation. As in the case of the free cooling gas studied in Ref.\ \cite{DG01}, I will use kinetic theory tools to explicitly get the dependence of $\epsilon$ on the restitution coefficients as well as on the mechanical parameters of the system (masses and sizes).  The theoretical estimates derived in this paper show that $\epsilon \neq 1$, although the deviations of the Einstein ratio from unity are in general quite small. This conclusion is in good agreement with the observations made by Barrat {\em et al.} \cite{BLP04}.

The paper is organized as follows. Section \ref{sec2} describes the problem I am interested in. In particular, two types of external forces (thermostats) to inject energy to the gas and reach a steady state are considered: the Gaussian and the stochastic thermostats. In the case of the Gaussian thermostat, the gas is heated by the action of an external force proportional to the velocity. This type of ``anti-drag'' force can be justified by Gauss's principle of least constraint and has been widely used in nonequilibrium molecular dynamics simulations of normal fluids \cite{EM90,H91}. Another mechanism of heating (similar to the one used in Ref.\ \cite{BLP04}) is to consider the action of a random external force, which gives frequent kicks to each particle between collisions. If this stochastic force has the properties of a white noise, it gives rise to a Fokker-Planck diffusion term in the Boltzmann  equation \cite{NE98}. Section \ref{sec3} is devoted to the calculation of the diffusion $D$ and mobility $\lambda$  coefficients by solving the Boltzmann-Lorentz equation by means of the Chapman-Enskog method \cite{CC70}. Both coefficients are explicitly determined by using both types of thermostats, being given in terms of the solutions of two linear integral equations. A practical evaluation of both transport coefficients is possible by using a Sonine polynomial expansion. Here, $D$ and $\lambda$ have been computed in the second Sonine approximation (namely, retaining two Sonine polynomials). As said above, in the case of the stochastic thermostat, the analysis agrees with the results of Barrat {\em et al.} \cite{BLP04}. In the case of the Gaussian thermostat, the deviations of $\epsilon$ from unity are more significant, although they are still small for not too inelastic systems. The paper is closed in Section \ref{sec4} with a brief discussion on the results obtained in this paper.

\section{Description of the problem}
\label{sec2}

Let us consider a granular gas of smooth inelastic hard spheres (of mass $m$, diameter $\sigma$, and interparticle coefficient of restitution $\alpha$)  in a homogeneous state. In the low-density regime, its velocity distribution function $f({\bf v})$ obeys the nonlinear Boltzmann equation \cite{BDS97}.  Due to dissipation in collisions, the gas is cooling unless a mechanism of energy input is externally introduced to compensate for collisional cooling. In experiments the energy is typically injected through the boundaries yielding a inhomogeneous steady state. To avoid the complication of strong temperature heterogeneities, it is usual to consider the action of  homogeneous external (driving) forces acting locally on each particle. These forces are called {\em thermostats} and depend on the state of the system. In this situation, the steady-state Boltzmann equation reads
\begin{equation}
\label{0}
{\cal F}f({\bf v})=J[{\bf v}|f,f],
\end{equation}
where the Boltzmann collision operator $J[{\bf v}|f,f]$ is 
\begin{equation}
J\left[{\bf v}_{1}|f,f\right]  =
\sigma^{2}\int d{\bf v}_{2}\int d\widehat{\boldsymbol{\sigma }}\,\Theta 
(\widehat{\boldsymbol{\sigma}}\cdot {\bf g}_{12})(\widehat{\boldsymbol {\sigma }}\cdot 
{\bf g}_{12}) 
 \left[ \alpha^{-2}f({\bf v}_{1}')f({\bf v}_{2}')-f({\bf v}_{1})f({\bf v}_{2})\right] 
\;. 
 \label{0.1}
\end{equation}
In Eq.\ (\ref{0}), ${\cal F}$ is an operator representing the effect of the external force,  $\widehat{\boldsymbol {\sigma}}$ is a unit vector 
along the line of centers of the two colliding particles, $\Theta$ is
the Heaviside step function, and ${\bf g}_{12}={\bf v}_{1}-{\bf v}_{2}$. In 
addition, the primes on the velocities denote the initial values $\{{\bf 
v}_{1}^{\prime},
{\bf v}_{2}^{\prime}\}$ that lead to $\{{\bf v}_{1},{\bf v}_{2}\}$
following a binary collision: 
\begin{equation}
\label{0.2}
{\bf v}_{1}^{\prime }={\bf v}_{1}-\case{1}{2}\left( 1+\alpha    
^{-1}\right)(\widehat{\boldsymbol {\sigma}}\cdot {\bf g}_{12})\widehat{\boldsymbol
{\sigma}},
\quad {\bf v}_{2}^{\prime}={\bf v}_{2}+\case{1}{2}\left( 
1+\alpha^{-1}\right) (\widehat{\boldsymbol {\sigma}}\cdot {\bf 
g}_{12})\widehat{\boldsymbol{\sigma}}\;.  
\end{equation}

Several types of thermostats can be used. As said in the Introduction, here I will consider two simple choices. One of them is a deterministic thermostat  based on Gauss's principle of least constraints \cite{EM90,H91}. In this case, ${\cal F}$ has the form 
\begin{equation}
\label{1}
{\cal F}f({\bf v})=\frac{1}{2}\zeta \frac{\partial}{\partial {\bf v}}\cdot \left[{\bf v}f({\bf v})\right],
\end{equation}
where $\zeta$ is the cooling rate due to collisions
\begin{equation}
\label{2}
\zeta=-\frac{1}{3nT}\int d{\bf v} mv^2 J[{\bf v}|f,f].
\end{equation}
Here, $m$ is the mass of a particle and $T$ is the granular temperature defined as 
\begin{equation}
\label{0.3}
\frac{3}{2}nT=\int\, d{\bf v}\, \frac{m}{2}v^2 f({\bf v}),
\end{equation}
$n$ being the number density
\begin{equation}
\label{0.3.1}
n=\int\, d{\bf v} f({\bf v}).
\end{equation}  
Another way of heating the gas is by the action of  a random force in the form of an uncorrelated white noise \cite{WK96}. The corresponding operator ${\cal F}$ has a Fokker-Planck form \cite{NE98}
\begin{equation}
\label{3}
{\cal F}f({\bf v})=-\frac{1}{2}\frac{T}{m}\zeta \left(\frac{\partial}{\partial {\bf v}}\right)^2f({\bf v}).
\end{equation}
The use of this stochastic force has attracted the attention of many theorists in the past years to study different problems, such as non-Gaussian properties (cumulants, high-energy tails) \cite{NE98,MS00}, long-range correlations \cite{NETP99}, collisional statistics and short-scale structure \cite{PTNE02}, transport properties \cite{GM02}, and fluctuation-dissipation relations \cite{PBL02}. 
\begin{figure}
\includegraphics[width=0.35 \columnwidth,angle=-90]{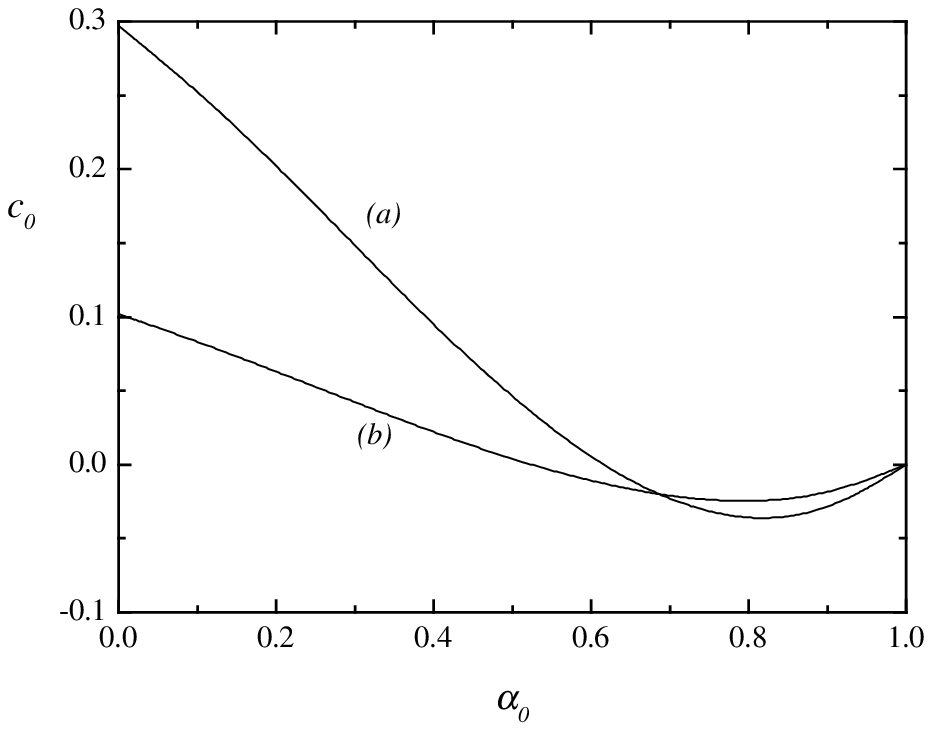}
\caption{Plot of the coefficient $c_0$ versus the coefficient of restitution $\alpha=\alpha_0$ for $m_0/m=0.5$ and $\sigma_0/\sigma=1$ for the Gaussian thermostat (a) and the stochastic thermostat (b). 
\label{fig1}}
\end{figure}

So far, the exact solution to the Boltzmann equation (\ref{0}) is not known, although a good approximation for thermal velocities can be obtained from an expansion in Sonine polynomials. In the leading order, $f$ is given by 
\begin{equation}
\label{5}
f({\bf v})\to n\pi^{-3/2}v_{\text{th}}^{-3}e^{-v^{*2}}\left[1+\frac{c}{4}\left(v^{*4}-5v^{*2}+\frac{15}{4}\right)\right],
\end{equation}
where $v^*=v/v_{\text{th}}$, $v_{\text{th}}=\sqrt{2T/m}$ being the thermal velocity. The coefficient $c$, which measures the deviation of $f$ from the Maxwellian form, is related to the kurtosis of the distribution.  Its value depends on the thermostat used and has been estimated from the Boltzmann equation up to linear order in $c$. In the case of the Gaussian thermostat (\ref{1}), the coefficient $c$ is given by \cite{NE98}    
\begin{equation}
\label{6}
c(\alpha)=\frac{32(1-\alpha)(1-2\alpha^2)}{81-17\alpha+30\alpha^2(1-\alpha)}.
\end{equation}
Estimate (\ref{6}) presents quite a good agreement with Monte Carlo simulations of the Boltzmann equation \cite{BMC96,MS00}. It is interesting to remark that in the homogenous problem the results obtained with and without the Gaussian thermostat are completely equivalent when one scales the particle velocity with respect to the thermal velocity \cite{MS00}. As a consequence, result (\ref{6}) applies to the free cooling case as well. 
In the case of the stochastic thermostat (\ref{3}) the result is \cite{NE98} 
\begin{equation}
\label{7}
c(\alpha)=\frac{32(1-\alpha)(1-2\alpha^2)}{241-177\alpha+30\alpha^2(1-\alpha)},
\end{equation}
As in the Gaussian case, Eq.\ (\ref{7}) agrees quite well with computer simulations \cite{MS00}. In general, both theoretical and numerical calculations show that, for not too inelastic systems, $c$ is very small. Deviations from the Maxwellian form are important, however, if one considers the tails of the distribution function, which are exponential for the stochastic thermostat and a ``stretched'' Gaussian for the Gaussian thermostat \cite{NE98,EP97,BCM99}. 
The cooling rate $\zeta$ can also be determined from the Sonine approximation (\ref{5}) with the result
\begin{equation}
\label{8}
\zeta=\frac{2}{3}\sqrt{2\pi}n \sigma^2 v_{\text{th}} (1-\alpha^2)\left(1+\frac{3}{32}c\right).
\end{equation}

We assume now that a few impurities (of mass $m_0$ and diameter $\sigma_0$) are added to the system. Given that their molar fraction is negligible, the state of the granular gas is not disturbed by the presence of impurities and so the velocity distribution function $f({\bf v})$ obeys the Boltzmann equation (\ref{0}). Moreover, one can also neglect collisions among impurities themselves versus the impurity-gas collisions, which are characterized by a coefficient of restitution $\alpha_0$. We want to analyze the diffusion of impurities immersed in the granular gas when the current of impurities is generated by the presence of a weak concentration gradient $\nabla n_0$ and/or a weak external field ${\bf E}$ (e.g. gravity or an electric field) acting only on the impurity particles.  Under these conditions, the velocity distribution function $f_0({\bf r}, {\bf v},t)$ of impurities verifies the Boltzmann-Lorentz equation 
\begin{equation}
\partial_t f_0+{\bf v}\cdot \nabla f_0+\frac{{\bf E}}{m_0}\cdot \frac{\partial}{\partial {\bf v}}f_0+{\cal F}f_0=J[{\bf v}|f_0,f],
\label{9}
\end{equation}  
where 
\begin{eqnarray}
J\left[{\bf v}_{1}|f_0,f\right] & =&
\overline{\sigma}^{2}\int d{\bf v}_{2}\int d\widehat{\boldsymbol{\sigma }}\,\Theta 
(\widehat{\boldsymbol{\sigma}}\cdot {\bf g}_{12})(\widehat{\boldsymbol {\sigma }}\cdot 
{\bf g}_{12}) \nonumber\\
& & \times
 \left[ \alpha_0^{-2}f_0({\bf r},{\bf v}_{1}';t)f({\bf r},{\bf v}_{2}';t)-f_0({\bf r},{\bf v}_{1};t)
f({\bf r}, {\bf v}_{2};t)\right] 
\;,
 \label{0.4}
\end{eqnarray}
$\overline{\sigma}=(\sigma+\sigma_0)/2$, and 
\begin{equation}
\label{0.5}
{\bf v}_{1}^{\prime }={\bf v}_{1}-\case{m}{m+m_0}\left( 1+\alpha_0    
^{-1}\right)(\widehat{\boldsymbol {\sigma}}\cdot {\bf g}_{12})\widehat{\boldsymbol
{\sigma}},
\quad {\bf v}_{2}^{\prime}={\bf v}_{2}+\case{m_0}{m+m_0}\left( 
1+\alpha_0^{-1}\right) (\widehat{\boldsymbol {\sigma}}\cdot {\bf 
g}_{12})\widehat{\boldsymbol{\sigma}}\;.  
\end{equation}
\begin{figure}
\includegraphics[width=0.35 \columnwidth,angle=-90]{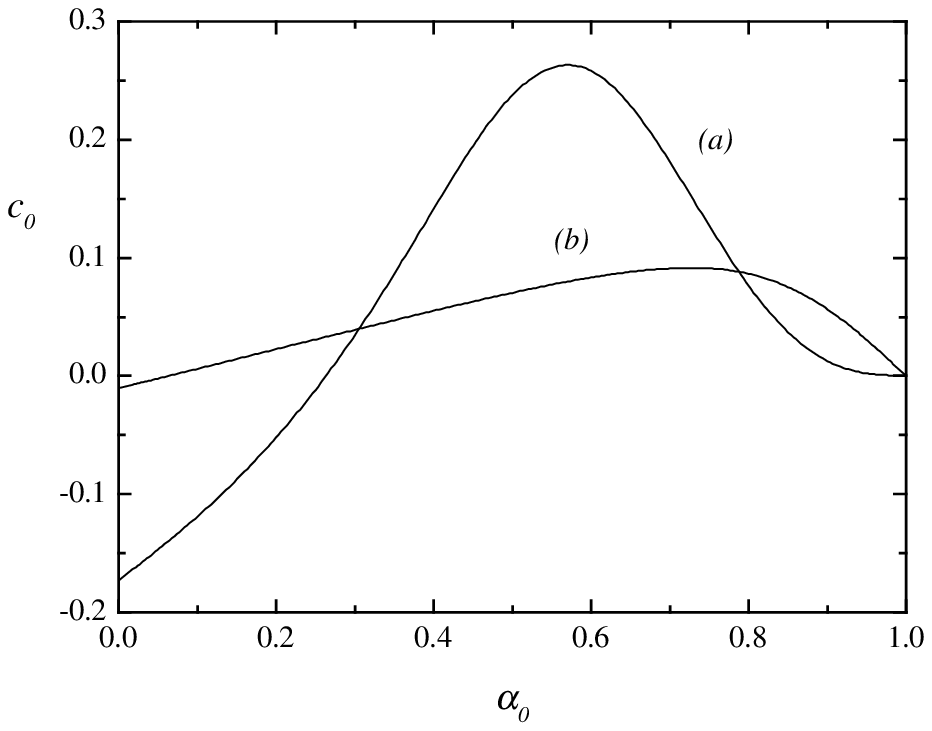}
\caption{Plot of the coefficient $c_0$ versus the coefficient of restitution $\alpha=\alpha_0$ for $m_0/m=5$ and $\sigma_0/\sigma=1$ for the Gaussian thermostat (a) and the stochastic thermostat (b). 
\label{fig1bis}}
\end{figure}
The partial temperature of impurities $T_0$ is defined as  
\begin{equation}
\label{0.6}
\frac{3}{2}n_0T_0=\int\, d{\bf v}\, \frac{m_0}{2}v^2 f_0({\bf v}),
\end{equation}
where $n_0$ is the number density of impurities,
\begin{equation}
\label{0.3.1bis}
n_0=\int\, d{\bf v} f_0({\bf v}).
\end{equation} 
 In addition, it is convenient to introduce the cooling rate $\zeta_0$ associated with the partial temperature $T_0$ of impurities: 
\begin{equation}
\label{10}
\zeta_0=-\frac{1}{3n_0T_0}\int d{\bf v} m_0v^2 J[{\bf v}|f_0,f].
\end{equation}

Momentum and energy are not collisional invariants of the Boltzmann-Lorentz collision operator $J[f_0,f]$. Only the number density of impurities is conserved: 
\begin{equation}
\label{0.7}
\partial_t n_0+\nabla\cdot {\bf j}_0=0, 
\end{equation}
 where the current of impurities ${\bf j}_0$ is defined as  
\begin{equation}
\label{0.8}
{\bf j}_0=\int d{\bf v} {\bf v}f_0({\bf v}).
\end{equation}
The conservation equation (\ref{0.7}) becomes a closed hydrodynamic equation for $n_0$ once ${\bf j}_0$ is expressed as a functional of $n_0$ and the external field ${\bf E}$. 

\begin{figure}
\includegraphics[width=0.35 \columnwidth,angle=-90]{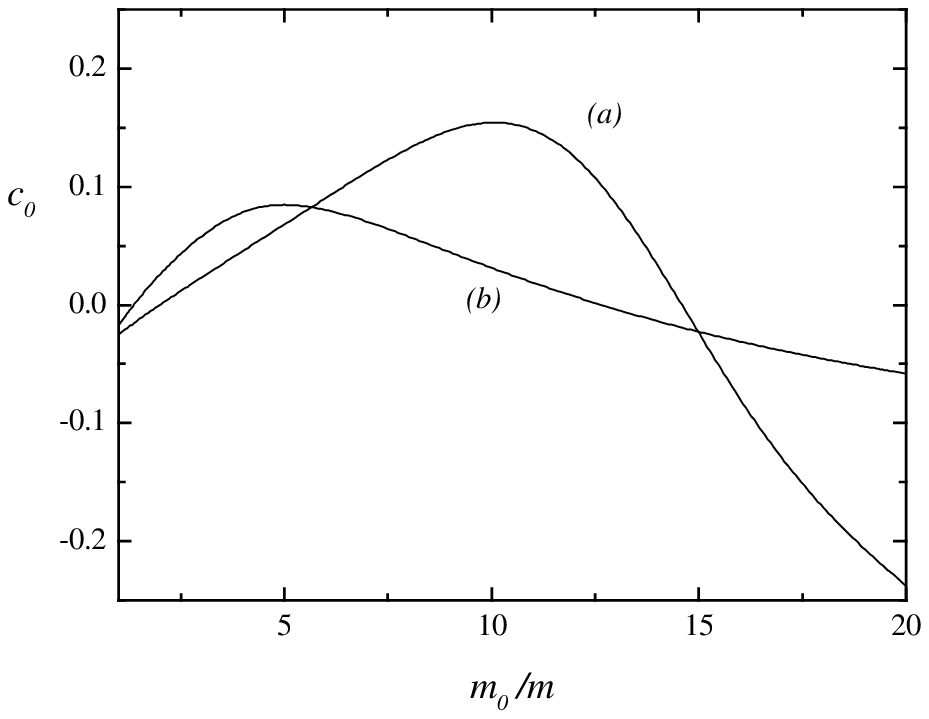}
\caption{Plot of the coefficient $c_0$ versus the mass ratio $m_0/m$ for $\alpha=\alpha_0=0.8$ and $\sigma_0/\sigma=1$ for the Gaussian thermostat (a) and the stochastic thermostat (b). 
\label{fig0}}
\end{figure}

In the absence of diffusion (homogeneous steady state), Eq.\ (\ref{9}) becomes 
\begin{equation}
\label{10.2}
{\cal F}f_0=J[{\bf v}|f_0,f].
\end{equation}
This equation has been recently analyzed by using both types of thermostats \cite{DHGD02,GM04}. The results show that the temperatures of the gas and impurities are clearly different ($T\neq T_0$) and so, the energy equipartition is broken down. The failure of the energy equipartition in granular gases has been confirmed by computer simulations \cite{DHGD02,MG02,simulations} and even observed in real experiments of vibrated mixtures in two \cite{WP02} and three \cite{FM02} dimensions. In general, the temperature ratio  $\gamma\equiv T_0/T$ presents a complex dependence on the parameters of the problem. However, according to Eqs.\ (\ref{1}),(\ref{3}), and (\ref{10.2}) the condition for determining the temperature ratio is different for each type of thermostat. In the case of the Gaussian thermostat (\ref{1}), when one multiplies both sides of Eq.\ (\ref{10.2}) by $v^2$ and integrates over ${\bf v}$, the steady state condition yields the equality of the cooling rates \cite{GM04,GD99}  
\begin{equation}
\label{10.2.1}
\zeta=\zeta_0.
\end{equation} 
In the case of the stochastic thermostat (\ref{3}), $\gamma$ is obtained from the condition \cite{DHGD02}
\begin{equation}
\label{10.2.2}
m_0\zeta=m \gamma \zeta_0. 
\end{equation} 
Requirements (\ref{10.2.1}) and (\ref{10.2.2}) lead to a different dependence of the temperature ratio $T_0/T$ on the control parameters, namely, the mass ratio $m_0/m$ , the size ratio  $\sigma_0/\sigma$ and the coefficients of restitution $\alpha$ and $\alpha_0$.  

\begin{figure}
\includegraphics[width=0.35 \columnwidth,angle=-90]{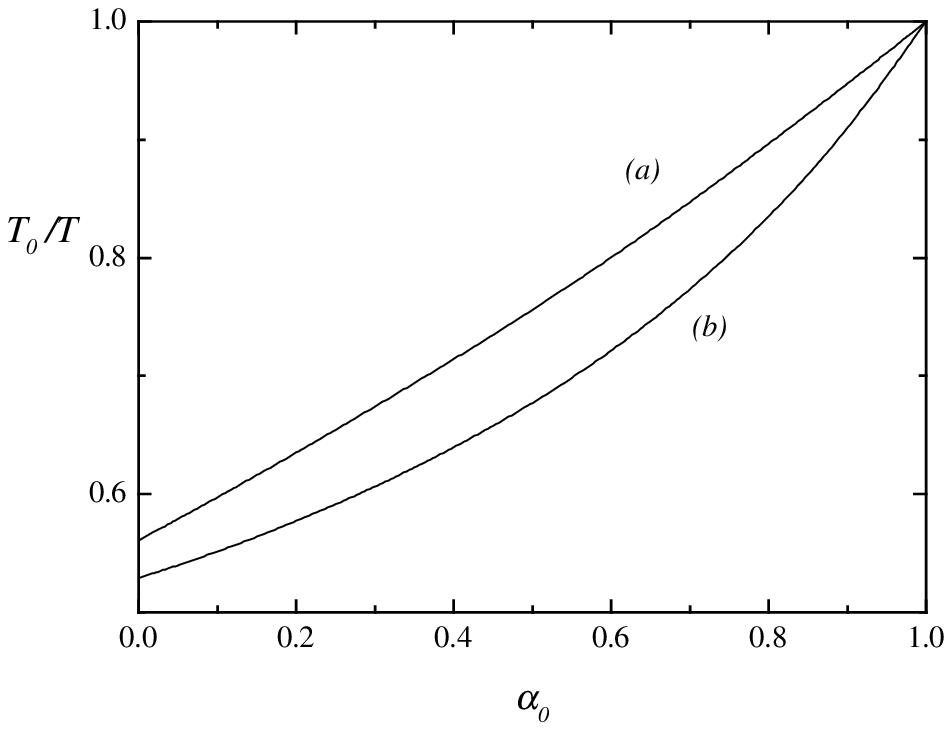}
\caption{Plot of the temperature ratio  $T_0/T$ versus the coefficient of restitution $\alpha=\alpha_0$ for $m_0/m=0.5$ and $\sigma_0/\sigma=1$ for the Gaussian thermostat (a) and the stochastic thermostat (b). 
\label{fig2}}
\end{figure}

A good estimate of $\gamma$ can be obtained by considering the first Sonine approximation to $f_0$:  
\begin{equation}
\label{10.1}
f_0({\bf v})\to n_0\pi^{-3/2}v_{\text{th}}^{-3}\theta^{3/2}e^{-\theta v^{*2}}\left[1+\frac{c_0}{4}\left(\theta^2v^{*4}-5\theta v^{*2}+\frac{15}{4}\right)\right],
\end{equation}
where $\theta=m_0T/mT_0$ is the mean square velocity of the gas particles relative to that of impurities. The coefficient $c_0$ can be determined by substitution of Eqs.\ (\ref{5}) and (\ref{10.1}) into the Boltzmann-Lorentz equation (\ref{10.2}) and retaining all terms linear in $c_0$ and $c$. The explicit expression of the coefficient $c_0$ is given in Appendix \ref{appA}. Once the coefficient $c_0$ is known, the temperature ratio can be obtained from the constraint  (\ref{10.2.1}) for the Gaussian thermostat  and (\ref{10.2.2}) for the stochastic thermostat.

Before illustrating the dependence of $c_0$ and $\gamma$ on the coefficients of restitution and the parameters of the mixture, it is instructive to consider some special limit cases.  First, when impurities are mechanically equivalent to the gas particles ($m=m_0$, $\sigma=\sigma_0$, $\alpha_0=\alpha$), the results of the single gas case \cite{NE98} are recovered, namely, $T=T_0$, and  
\begin{equation}
\label{a12}
c_0=\frac{32(1-\alpha_0)(1-2\alpha_0^2)}{81-17\alpha_0+30\alpha_0^2(1-\alpha_0)},
\end{equation}
for the Gaussian thermostat while 
\begin{equation}
\label{a13}
c_0=\frac{32(1-\alpha_0)(1-2\alpha_0^2)}{241-177\alpha_0+30\alpha_0^2(1-\alpha_0)},
\end{equation}
for the stochastic thermostat. Also, when the collisions among gas particles are elastic ($\alpha=1$ and so $c=0$), our Sonine solution gives $c_0=0$ and 
\begin{equation}
\label{10.2.3}
\gamma=\frac{1+\alpha_0}{2+(1-\alpha_0)(m/m_0)}
\end{equation}
for both types of thermostats. This result agrees with the {\em exact} solution of the Boltzmann-Lorentz equation found by Martin and Piasecki \cite{MP99} in the study of the stationary state of a test particle immersed in a homogeneous fluid in equilibrium, undergoing dissipative collisions with the fluid particles. Finally, in the weak dissipation limit ($1-\alpha \ll 1$,  $1-\alpha_0 \ll 1$), the temperature ratio $\gamma$ behaves as 
\begin{equation}
\label{10.2.4}
\gamma \to 1-\frac{m+m_0}{2m_0}(1-\alpha_0)+
\left(\frac{\sigma}{\overline{\sigma}}\right)^2\frac{1}{2m}\sqrt{\frac{(m+m_0)^3}{2m_0}}(1-\alpha),
\end{equation}
for the Gaussian thermostat and 
\begin{equation}
\label{10.2.5}
\gamma \to 1-\frac{m+m_0}{2m_0}(1-\alpha_0)+
\left(\frac{\sigma}{\overline{\sigma}}\right)^2\frac{1}{2m^2}\sqrt{\frac{m_0(m+m_0)^3}{2}}(1-\alpha),
\end{equation}
for the stochastic thermostat. The two latter terms on the right hand side of Eqs.\  (\ref{10.2.4}) and (\ref{10.2.5}) represent two different types of inelastic collisions providing independent mechanisms to enforce the breakdown of energy equipartition.

\begin{figure}
\includegraphics[width=0.35 \columnwidth,angle=-90]{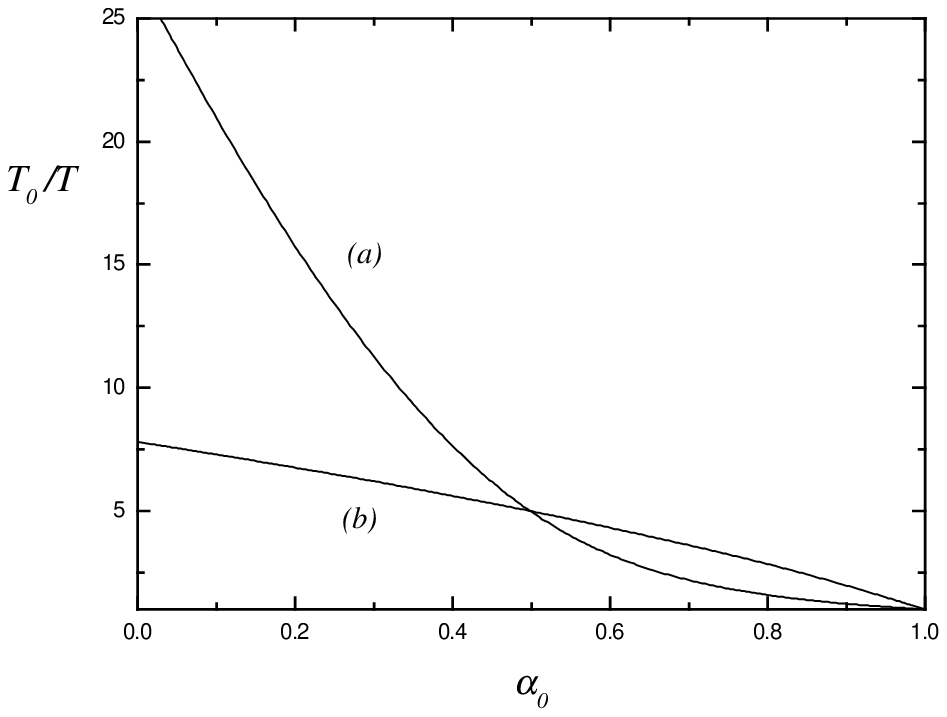}
\caption{Plot of the temperature ratio  $T_0/T$ versus the coefficient of restitution $\alpha=\alpha_0$ for $m_0/m=5$ and $\sigma_0/\sigma=1$ for the Gaussian thermostat (a) and the stochastic thermostat (b). 
\label{fig3}}
\end{figure}

Beyond the above limit cases, it is quite difficult to provide the explicit dependence of $c_0$ and $\gamma$ on the characteristic parameters of the system. For the sake of concreteness, now we consider the case $\alpha=\alpha_0$. Figures\ \ref{fig1} and \ref{fig1bis} show the coefficient $c_0$ as a function of $\alpha_0$ for $\sigma_0/\sigma=1$ and the mass ratios  $m_0/m=0.5$ and $m_0/m=5$, respectively. We consider the cases of the Gaussian and stochastic thermostats. When the mass ratio is smaller than 1, we see that the dependence of $c_0$ on $\alpha_0$ is qualitatively similar in both thermostats, its value being small for not too inelastic systems. This means that for moderate dissipation the distribution function $f_0$ of the homogeneous state is quite close to a Maxwellian at the temperature of the impurity particle $T_0$. However, according to Fig.\ \ref{fig1bis}, the dependence of $c_0$ on $\alpha_0$ is quantitatively different for both thermostats when impurities are heavier than the particles of the gas. In particular, the curves of Fig.\ \ref{fig1bis} exhibit a different convexity in the quasielastic limit. To know something more on the dependence of $c_0$ on the mass ratio, in Fig.\ \ref{fig0} we plot $c_0$ versus the mass ratio $m_0/m$  for $\alpha=\alpha_0=0.8$ and $\sigma_0/\sigma=1$. The qualitative dependence of $c_0$ on $m_0/m$ is quite similar for both thermostats since first $c_0$ increases with $m_0/m$, reaches a maximum and then decreases with the mass ratio. At a quantitative level, we observe that  the magnitude of $c_0$ for the stochastic thermostat is in general smaller than that of the Gaussian thermostat and consequently, the deviation of $f_0$ from its Maxwellian form is more significant in the case of the Gaussian thermostat than in the case of the stochastic force.
The dependence of the temperature ratio $T_0/T$ on the restitution coefficient $\alpha_0$ is illustrated in Figs.\ \ref{fig2} and \ref{fig3} for the cases $m_0/m=0.5$ and $\sigma_0/\sigma=1$ and $m_0/m=5$ and $\sigma_0/\sigma=1$, respectively. We consider again both types of thermostats.  We observe that the temperature of the impurity is smaller (larger) than that of the gas when the impurity is lighter (heavier) than the particles of the gas, whatever the thermostat considered is. The lack of energy equipartition is quite apparent in all the cases studied, especially when the mass ratio $m_0/m$ is larger than 1 in the case of the Gaussian thermostat. This is indicative of what happens in the homogeneous cooling state in the large impurity/gas mass ratio where a peculiar ``phase transition'' has been recently found for which the diffusion coefficient is normal in one phase but grows without bound in the other \cite{SD01}.  
It must be remarked that the predictions obtained for $c_0$ and $\gamma$ from the leading Sonine approach (\ref{10.1}) compare quite well with recent computer simulations \cite{MG02,GM04}. 

\section{Diffusion and mobility coefficients}
\label{sec3}

As said in the Introduction, my aim is to get the diffusion and mobility coefficients in the limit of small concentration gradient $\nabla n_0$ and weak external fields. These transport coefficients are determined by solving the Boltzmann-Lorentz equation (\ref{9}) from the Chapman-Enskog method \cite{CC70}. This method assumes the existence of a {\em normal} solution  in which all the space and time dependence of $f_0$ is through the hydrodynamic fields. In this problem, the normal solution is explicitly generated by expanding $f_0$ in powers of the gradient $\nabla n_0$ and the field strength ${\bf E}$:
\begin{equation}
\label{11}
f_0=f_0^{(0)}+\delta\, f_0^{(1)}+\cdots, 
\end{equation}
where each factor $\delta$ corresponds to the implicit factors $\nabla n_0$ and ${\bf E}$.  The time derivative is also expanded as $\partial_t=\partial_t^{(0)}+\delta\, \partial_t^{(1)}+\cdots$, where the action of the different operators $\partial_t^{(k)}$ can be obtained from the balance equation (\ref{0.7}). They are given by 
\begin{equation}
\label{12}
\partial_t^{(0)}n_0=0,\quad \partial_t^{(k)}n_0=-\nabla \cdot {\bf j}_0^{(k-1)},\quad  k \geq 1,
\end{equation}
where  
\begin{equation}
\label{14}
{\bf j}_0^{(k)}=\int d{\bf v} {\bf v}f_0^{(k)}.
\end{equation}

The zeroth-order approximation $f_0^{(0)}$ is the solution of Eq.\ (\ref{10.2}), whose approximate form is given by Eq.\ (\ref{10.1})  but taking into account now the local dependence on the density $n_0$. Since $f_0^{(0)}$ is isotropic, it follows that the flux of impurities vanishes at this order, i.e., ${\bf j}_0^{(0)}={\bf 0}$, and so $\partial_t^{(1)}n_0=0$. To first order in $\delta$, one has the kinetic equation
\begin{eqnarray}
\label{15}
{\cal F}f_0^{(1)}-J[{\bf v}|f_{0}^{(1)},f]&=&-\left({\bf v}\cdot \nabla +\frac{{\bf E}}{m_0}\cdot \frac{\partial}{\partial {\bf v}}\right)f_0^{(0)}\nonumber\\
&=& -\left[{\bf v}\cdot \left(\nabla \ln n_0\right)+\frac{{\bf E}}{m_0}\cdot \frac{\partial}{\partial {\bf v}}\right]f_0^{(0)}.
\end{eqnarray} 
The second equality follows from the space dependence of $f_0^{(0)}$ through $n_0$. The solution to Eq.\ (\ref{15}) has the form
\begin{equation} 
\label{16}
f_0^{(1)}({\bf v})={\boldsymbol {\cal A}}({\bf v})\cdot \nabla \ln n_0+{\boldsymbol {\cal B}}({\bf v})\cdot {\bf E}.
\end{equation}
Substitution of Eq.\ (\ref{16}) into Eq.\ (\ref{15}) leads to the set of linear integral equations
\begin{equation}
\label{17}
{\cal F}{\boldsymbol {\cal A}}-J[{\bf v}|{\boldsymbol {\cal A}},f]=-{\bf v} f_0^{(0)},
\end{equation}
\begin{equation}
\label{18}
{\cal F}{\boldsymbol {\cal B}}-J[{\bf v}|{\boldsymbol {\cal B}},f]=-\frac{1}{m_0}\frac{\partial}{\partial {\bf v}}f_0^{(0)}.
\end{equation}
To first order, the current has the structure 
\begin{equation}
\label{19}
{\bf j}_0^{(1)}=-D \nabla \ln n_0+\lambda {\bf E},
\end{equation}
where $D$ is the diffusion coefficient and $\lambda$ is the mobility coefficient.  According to Eqs.\  (\ref{14}) and (\ref{16}),  these coefficients are given by 
\begin{equation}
\label{20}
D=-\frac{1}{3}\int d{\bf v}{\bf v} \cdot {\boldsymbol {\cal A}},
\end{equation}
\begin{equation}
\label{21}
\lambda=\frac{1}{3}\int d{\bf v}{\bf v} \cdot {\boldsymbol {\cal B}}.
\end{equation}

In the elastic limit ($\alpha=\alpha_0=1$), $T=T_0$, and so $f_0^{(0)}$ is the local equilibrium distribution, i.e.,
\begin{equation}
\label{0.9}
f_0^{(0)}({\bf v})=n_0\left(\frac{m_0}{2\pi T}\right)^{3/2}\exp\left(-\frac{m_0v^2}{2T}\right).
\end{equation}
In this case, $\partial f_0^{(0)}/\partial {\bf v}=-(m_0{\bf v}/T)f_0^{(0)}$ and the integral equations (\ref{17}) and (\ref{18}) yield ${\boldsymbol {\cal A}}=-T{\boldsymbol {\cal B}}$. As a consequence, one recovers the usual Einstein relation between the diffusion and mobility coefficients
\begin{equation}
\label{0.10}
\epsilon=\frac{D}{T\lambda}=1.
\end{equation}

However, at finite inelasticity the relationship between $D$ and $\lambda$ is no longer simple. In the driven case, there are two separate reasons for which the Einstein relation in its usual form is violated. First, since $f_0^{(0)}$ is not a Gaussian distribution then,  $\partial f_0^{(0)}/\partial {\bf v}
 \neq -(m_0{\bf v}/T)f_0^{(0)}$ and $D$ is not proportional to $\lambda$. The second reason is that the temperature of the impurity particle is different from that of the fluid, except when the impurity is mechanically equivalent to particles of the fluid. This last source of discrepancy could be perhaps eliminated if one replaces the temperature of the gas $T$ by the temperature of impurities $T_0$ in the usual Einstein ratio (\ref{0.10}). This change leads to the modified Einstein ratio given by Eq.\ (\ref{n1}). 
As a matter of fact, if one took the Maxwellian approximation (\ref{0.9}) for $f_0^{(0)}$ (with $T_0$ instead of $T$), then $\partial f_0^{(0)}/\partial {\bf v}
=-(m_0{\bf v}/T_0)f_0^{(0)}$ and  the modified Einstein relation (\ref{n1}) between the diffusion and mobility coefficients would apply ($\epsilon=1$). Consequently, the only reason for which $\epsilon \neq 1$ is due to the non-Maxwellian behavior of the velocity distribution function $f_0^{(0)}$ (absence of the Gibbs state), which leads to small deviations that could be difficult to detect in computer simulation experiments.  This conclusion agrees with the results obtained from recent molecular dynamics simulations \cite{BLP04}, where no deviations from the Einstein relation $\epsilon=1$ were observed for a wide range of values of the coefficients of restitution and parameters of the system.  

For practical purposes, the linear integral equations (\ref{17}) and (\ref{18}) can be solved by using the first terms in a Sonine polynomial expansion. The reliability of this Sonine approximation has been confirmed in several {\em inhomogeneous} situations. Thus, in the case of the shear viscosity coefficient for a single gas, the first Sonine predictions \cite{GM02,BMC99} compare quite well with a numerical solution of the Boltzmann equation in the uniform shear flow state obtained from the direct simulation Monte Carlo (DSMC) method \cite{B94}. Such a good agreement for the shear viscosity has been also extended to the case of multicomponent systems in the low-density limit \cite{MG03} and for finite densities \cite{GM03}. More recently, the tracer diffusion coefficient has been evaluated up to the second order of the Sonine expansion \cite{GM04}. As in the case of the shear viscosity coefficient, the Sonine solution agrees quite well with computer simulation results over a wide region of the parameter space. In this paper, the diffusion and mobility coefficients will be determined up to the second Sonine approximation. In this case, the unknowns ${\boldsymbol {\cal A}}$ and ${\boldsymbol {\cal B}}$ are approximated by 
\begin{equation}
\label{23}
{\boldsymbol {\cal A}}({\bf v})\to f_{0,M}({\bf v})\left[a_{1}{\bf v}+a_{2}{\bf S}_0({\bf v})
\right], 
\end{equation}
\begin{equation}
\label{23.1}
{\boldsymbol {\cal B}}({\bf v})\to f_{0,M}({\bf v})\left[b_{1}{\bf v}+b_{2}{\bf S}_0({\bf v})
\right], 
\end{equation}
where $f_{0,M}({\bf v})$ is a Maxwellian distribution at the temperature $T_0$ of the impurities, i.e.
\begin{equation}
\label{24}
f_{0,M}({\bf v})=n_0\left(\frac{m_0}{2\pi T_0}\right)^{3/2}\exp\left(-\frac{m_0v^2}{2T_0}\right),
\end{equation}
and ${\bf S}_0({\bf v})$ is the polynomial 
\begin{equation}
\label{25}
{\bf S}_0({\bf v})=\left(\frac{1}{2}m_0v^2-\frac{5}{2}T_0\right){\bf v}.
\end{equation}
The coefficients $a_{1}$, $b_{1}$, $a_2$, and $b_2$  are defined as 
\begin{equation}
\label{26}
\left(
\begin{array}{c}
a_{1}\\
b_1
\end{array}
\right)
=\frac{m_0}{3n_0T_0}\int d{\bf v}\,{\bf v} \cdot 
\left(
\begin{array}{c}
{\boldsymbol {\cal A}}\\
{\boldsymbol {\cal B}}
\end{array}
\right),
\end{equation}
\begin{equation}
\label{27}
\left(
\begin{array}{c}
a_{2}\\
b_2
\end{array}
\right)
=\frac{2}{15}\frac{m_0}{n_0T_0^3}\int d{\bf v}\,{\bf S}_0 
  \cdot 
\left(
\begin{array}{c}
{\boldsymbol {\cal A}}\\
{\boldsymbol {\cal B}}
\end{array}
\right).
\end{equation}
These coefficients are determined by substitution of Eqs.\ (\ref{23}) and (\ref{23.1}) into the integral equations (\ref{17}) and (\ref{18}), respectively. The details are carried out in Appendix \ref{appB}. In the case of the Gaussian thermostat, the Einstein ratio (\ref{n1}) becomes 
\begin{equation}
\label{28}
\epsilon=1-\frac{c_0}{2}\frac{\nu_2^*}{\nu_4^*-\frac{3}{2}\zeta^*},
\end{equation}
where $\zeta^*$, $\nu_2^*$ and $\nu_4^*$ are explicitly given in Appendix \ref{appB}. In the case of the stochastic thermostat, one gets
\begin{equation}
\label{29}
\epsilon=1-\frac{c_0}{2}\frac{\nu_2^*}{\nu_4^*}.
\end{equation}
   
\begin{table}[tbp]
\caption{Values of the Einstein ratio $\epsilon$ for the different cases studied in Ref.\ \cite{BLP04}. The last two columns refer to the  results obtained by using the Gaussian and stochastic thermostats, respectively.} 
\label{table1}
\begin{tabular}{ccccc}
\\ 
\hline
\hline
$m_0/m$&\quad $\alpha$&\quad $\alpha_0$&\quad $\epsilon$&\quad $\epsilon$ \\ \hline
1&\quad 0.9&\quad 0.4&\quad 1.002& \quad 1.001\\ 
4&\quad 0.7&\quad 0.7&\quad 0.991&\quad 0.997\\
7&\quad 0.7&\quad 0.7& \quad 0.995& \quad 0.998 \\
\hline
\hline
\end{tabular}
\end{table}

\vspace{0.5cm}

Equations (\ref{28}) and (\ref{29}) clearly show that the deviations from the Einstein relation $\epsilon=1$ are proportional to the coefficient $c_0$, which measures the deviation of the velocity distribution function of impurities from the Maxwellian distribution. Both $c$ and $c_0$ vanish in the limit of elastic collisions and otherwise give contributions of the order of a few percent to $\nu_2^*$, $\nu_4^*$, and $\zeta^*$, even at very strong dissipation. According to Eqs.\ (\ref{28}) and (\ref{29}) it is apparent that the Einstein ratio exhibits a complex dependence on $\alpha$, $\alpha_0$, and the mechanical parameters of the mixture (the mass and size ratios). A simpler form for the Einstein ratio corresponds to the case of mechanically equivalent particles (i.e., $m=m_0$, $\sigma=\sigma_0$, and $\alpha=\alpha_0$). In this case, $T_0=T$, $c_0=c$, and Eqs.\ (\ref{28}) and (\ref{29}) reduce in the quasielastic limit to a common ratio given by 
\begin{equation}
\label{30}
\epsilon\to 1+\frac{5}{118}(1-\alpha).
\end{equation}  
Equation (\ref{30}) shows most directly the dependence of $\epsilon$ on dissipation. Also, when the gas is at equilibrium ($\alpha=1$), then $c_0=0$ with $\gamma$ given by Eq.\ (\ref{10.2.3}) and so, the Einstein relation is exactly verified.

\begin{figure}
\includegraphics[width=0.35 \columnwidth,angle=-90]{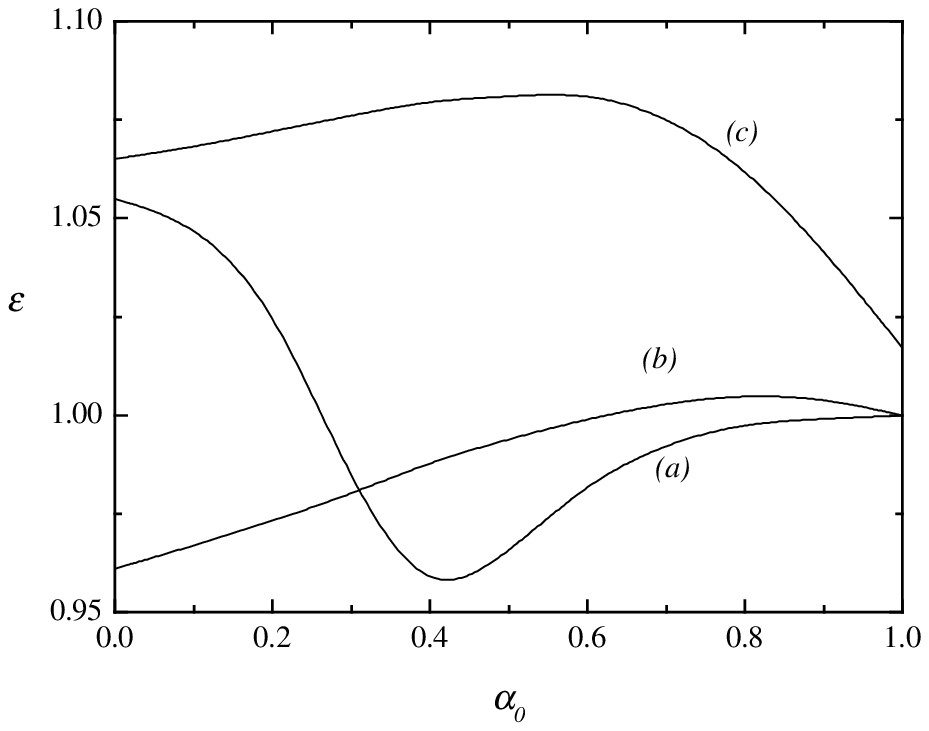}
\caption{Plot of the Einstein ratio $\epsilon$ versus the coefficient of restitution $\alpha_0$ for the Gaussian thermostat in the cases: (a) $\alpha=\alpha_0$, $m_0/m=5$ and $\sigma_0/\sigma=1$; (b)  $\alpha=\alpha_0$, $m_0/m=0.5$ and $\sigma_0/\sigma=1$; and (c) $\alpha=0.5$, $m_0/m=10$ and $\sigma_0/\sigma=1$.  
\label{fig4}}
\end{figure}

As said in the Introduction, very recently Barrat {\em et al.} \cite{BLP04} have used the DSMC method \cite{B94} and molecular dynamics simulations to investigate the validity of fluctuation-dissipation relations for a binary mixture of inelastic hard disks with identical diameters but different physical properties. The system is also subjected to a external force which has the form of a white noise [Eq.\ (\ref{3})] with the addition of a viscous term. In the single tracer case, their computer simulation results clearly support the Einstein relation (\ref{n1}). 
Although a quantitative comparison between the theory presented here and their results is not possible,  it is illustrative to get the values of the Einstein ratio $\epsilon$ predicted by the theory for the same cases as considered in the simulations. The results are displayed in Table \ref{table1} for the Gaussian and stochastic thermostats. Note that the latter coincides with the thermostat used in simulations, except for the viscous term. We see that the deviations from the Einstein relation are very small (less than 1\%), even for moderate dissipation (say for instance, $\alpha=\alpha_0=0.7$). This confirms the conclusions obtained in Ref.\ \cite{BLP04} from computer simulations.

To illustrate the influence of dissipation on the Einstein ratio more generally, in Figs.\ \ref{fig4} and \ref{fig5} the Einstein ratio $\epsilon$ is plotted versus $\alpha_0$ for $\sigma_0/\sigma=1$ and different values of the mass ratio $m_0/m$. Figure  \ref{fig4} shows the results obtained by using the Gaussian thermostat while Fig.\ \ref{fig5} refers to the results obtained from the stochastic thermostat. Although the mechanical properties between both species (impurities and particles of the gas) are quite different, we observe that $\epsilon$ is very close to 1, especially in the case of the stochastic thermostat where again the deviations are smaller than 1\%. However, in the case of the Gaussian thermostat the deviations from unity are even about 8\% (for instance, for $\alpha=\alpha_0=0.5$ and $m_0/m=10$). 
This failure of the Einstein relation could be detected in computer simulations by using  the Gaussian thermostat instead of stochastic driving. In addition, Figs.\  \ref{fig4} and \ref{fig5} illustrate again the fact that the transport properties are affected by the thermostat introduced so that the latter does not play a neutral role in the problem \cite{GM02}.

\section{Discussion}
\label{sec4}

The goal of this paper has been to analyze the validity of the Einstein relation between the diffusion and mobility coefficients in a heated granular gas. This analysis has been mainly motivated by recent computer simulations performed by Barrat {\em et al.} \cite{BLP04} where no apparent deviations of the Einstein ratio from unity were found when the temperature of the gas $T$ is replaced by the temperature of impurities $T_0$ in the usual Einstein ratio, Eq.\ (\ref{n1}). This conclusion contrasts with the one obtained before by Dufty and Garz\'o \cite{DG01} in the {\em unforced} gas case where it was clearly shown the violation of the Einstein relation. The analysis carried out in Ref.\ \cite{DG01} indicates that the deviations of the Einstein ratio in its usual form from unity has three distinct origins: non-Gaussianity of the distribution function of the homogeneous cooling state, energy nonequipartition, and time evolution of the granular temperature.  Given that the last source of discrepancy is no present in the driven case, it is interesting to assess the effect on the violation of the Einstein relation coming from the first two reasons. 

\begin{figure}
\includegraphics[width=0.35 \columnwidth,angle=-90]{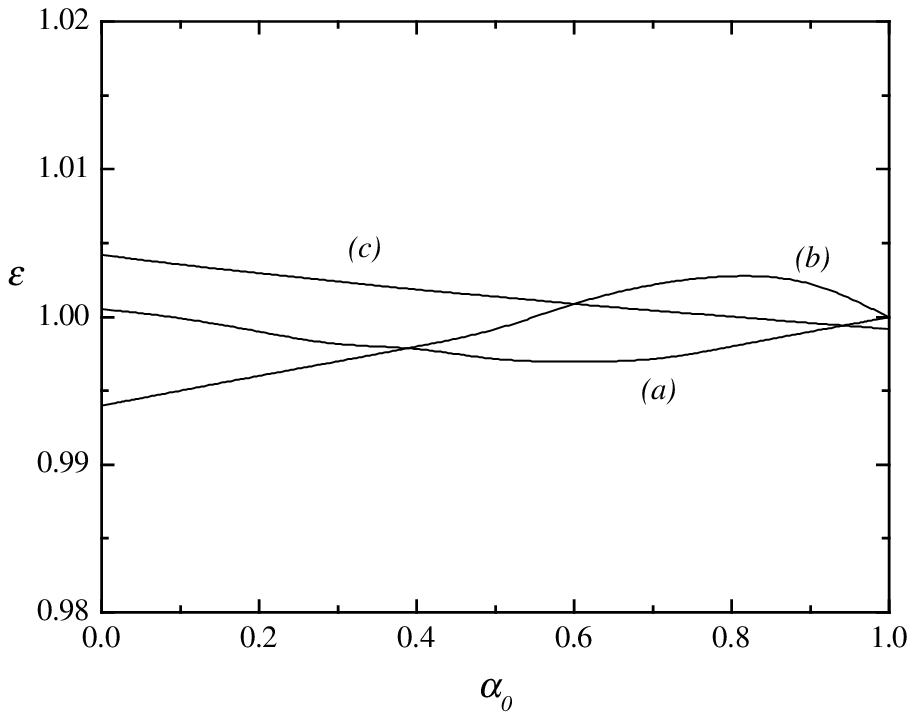}
\caption{Plot of the Einstein ratio $\epsilon$ versus the coefficient of restitution $\alpha_0$ for the stochastic thermostat in the cases: (a) $\alpha=\alpha_0$, $m_0/m=5$ and $\sigma_0/\sigma=1$; (b)  $\alpha=\alpha_0$, $m_0/m=0.5$ and $\sigma_0/\sigma=1$; and (c) $\alpha=0.5$, $m_0/m=10$ and $\sigma_0/\sigma=1$.  
\label{fig5}}
\end{figure}

To achieve a steady state, the granular gas is driven by the action of external forces (thermostats) that accelerate the particles and hence compensate for collisional cooling. Here, I have used two types of thermostats: (a) an ``anti-drag'' force proportional to the particle velocity (Gaussian force), and (b) a stochastic force, which gives frequent kicks to each particle between collisions. The introduction of these thermostats has the advantage of avoiding the intrinsic time dependence of the temperature in the free cooling state, but at the price of introducing new effects induced by the external forcing. The diffusion $D$ and mobility $\lambda$ coefficients have been determined by solving the  
Boltzmann-Lorentz equation by means of the Chapman-Enskog method. In the first order of expansion, $D$ and $\lambda$ are given in terms of the solutions of two linear integral equations. An approximate solution of these integral equations has been obtained by expanding the distribution function of impurities in a Sonine polynomial expansion. Here, I have retained two Sonine polynomials (second Sonine approximation). As happens for the transport properties of a single gas \cite{GM02}, the Sonine solution shows that the dependence of $D$ and $\lambda$ on the parameters of the problem is different for each type of thermostat. Although not widely recognized, this conclusion illustrates the fact that generally the inclusion of an external force depending on the state of the system changes the apparent transport coefficients.

As Barrat {\em et al.} \cite{BLP04} suggests, the influence of the temperature differences between impurities and particles of the gas on the Einstein ratio can be mitigated in part if one uses $T_0$ instead of $T$ in the usual Einstein ratio. For this reason,  I have explored the validity  of the modified Einstein ratio (\ref{n1}). The analytical results (based on the second Sonine approximation) show that the deviation of the Einstein ratio from unity is {\em only} due to the factor $c_0$ which measures the deviations of the velocity distribution function of the impurities from a Maxwellian.  In fact, if one neglects the small effect of this factor, then $\epsilon=1$.  In addition, the results provided here also show that the deviation of $\epsilon$ from 1 is more significant in the case of the Gaussian thermostat than in the case of the stochastic one.  This is due essentially to the fact that in general the magnitude of the coefficient $c_0$ of the stochastic thermostat is smaller than that of the Gaussian thermostat. 
As an illustration, Figs.\ \ref{fig4} and \ref{fig5} show that the deviations of the Einstein ratio from unity are generally  smaller than 1\% for the stochastic thermostat while these deviations are even about 8\% in the case of the Gaussian thermostat. 
Given that the simulations carried in Ref.\ \cite{BLP04} were made by using a stochastic driving (with the addition of a viscous term), the results of this paper agree with their observations since deviations smaller than 1\% 
are quite difficult to detect in computer simulations.

The results obtained here have been restricted to the single tracer case, i.e., a especial limit case of a binary mixture. Barrat {\em et al.}'s computer simulations also show the validity of the Einstein relations separately for each species, each one with its own temperature.  I plan to analyze this issue in the next future. 

\acknowledgments  
I acknowledge partial support from the Ministerio de Ciencia y Tecnolog\'{\i}a (Spain) through Grant No. FIS2004-01399.

\appendix
\section{Explicit expressions of $\zeta_0$ and $c_0$}
\label{appA} 
 
In this Appendix, the expressions of the partial cooling rate $\zeta_0$ and the coefficient $c_0$ in the homogenous states driven by the Gaussian and stochastic thermostats are given. Most of these results have been recently reported by the author in Ref.\ \cite{GM04} for the homogeneous cooling state. By using the leading Sonine approximations (\ref{5}) and (\ref{10.1}) and neglecting nonlinear terms in $c$ and $c_0$, $\zeta_0$ can be written as \cite{GD99,GM04}
\begin{equation}
\label{a1}
\zeta_0=\lambda_{00}+\lambda_{01}c_0+\lambda_{02}c,
\end{equation}
where 
\begin{equation}
\label{a2}
\lambda_{00}=\frac{8}{3}\sqrt{\pi}n\overline{\sigma}^2v_{\text{th}} \mu
\left(\frac{1+\theta}{\theta}\right)^{1/2}(1+\alpha_{0}) 
\left[1-\frac{\mu}{2}(1+\alpha_{0})(1+\theta)\right],
\end{equation} 
\begin{equation}
\label{a3}
\lambda_{01}=\frac{1}{12}\sqrt{\pi}n\overline{\sigma}^2v_{\text{th}} \mu
\frac{(1+\theta)^{-3/2}}{\theta^{1/2}}(1+\alpha_{0}) 
\left[2(3+4\theta)-3\mu(1+\alpha_{0})(1+\theta)
\right],
\end{equation}
\begin{equation}
\label{a4}
\lambda_{02}=-\frac{1}{12}\sqrt{\pi}n\overline{\sigma}^2v_{\text{th}} \mu
\left(\frac{1+\theta}{\theta}\right)^{-3/2}(1+\alpha_{0}) 
\left[2+3\mu(1+\alpha_{0})(1+\theta)
\right].
\end{equation}
Here, $\overline{\sigma}=(\sigma+\sigma_0)/2$, $\mu=m/(m+m_0)$, and $\theta=m_0T/mT_0$.

The coefficient $c_0$ is defined by 
\begin{equation}
\label{a5}
c_0=\frac{8}{15}\left(\frac{m_0^2}{4n_0T_0^2}\int\, d{\bf 
v}v^4f_0-\frac{15}{4}\right).
\end{equation}
This coefficient is determined from the Boltzmann-Lorentz equation (\ref{10.2}) by multiplying that equation by $v^4$, and integrating over the velocity. When only linear terms in $c_0$ and $c$ are retained, the result in the case of the Gaussian thermostat is found to be \cite{GD99}
\begin{equation}
\label{a6}
-\frac{15}{2}\frac{\mu_0^2}{\theta^2}\zeta_0\left(1+\frac{c_0}{2}\right)=\Omega _{00}+\Omega _{01}c_0+\Omega _{02}c,
\end{equation}
while in the case of the stochastic thermostat the result is 
\begin{equation}
\label{a6.1}
-\frac{15}{2}\frac{\mu_0^2}{\theta^2}\zeta_0=\Omega _{00}+\Omega _{01}c_0+\Omega _{02}c,
\end{equation}
where use has been made of the conditions (\ref{10.2.1}) and (\ref{10.2.2}), respectively.
Here, $\mu_0=1-\mu=m_0/(m+m_0)$ and I have introduced the quantities \cite{GM04}
\begin{eqnarray}
\label{a7}
\Omega _{00} &=&2\sqrt{\pi}n\overline{\sigma}^2v_{\text{th}} \mu_0^2\mu \frac{\left(
1+\theta \right) ^{-1/2}}{\theta^{5/2}}\left( 1+\alpha _{0}\right) 
\left[ -2\left( 6+5\theta\right) +\mu \left(1+\alpha_{0}\right) 
\left( 1+\theta \right) \left( 14+5\theta \right)\right.\nonumber\\
& &\left. 
 -8\mu^{2}\left( 1+\alpha _{0}\right) ^{2}\left( 1+\theta \right) ^{2} 
+2\mu^{3}\left( 1+\alpha _{0}\right) ^{3}\left(
1+\theta \right) ^{3}\right],  
\end{eqnarray}
\begin{eqnarray}
\label{a8}
\Omega _{01} &= & \frac{\sqrt{\pi}}{8}n\overline{\sigma}^2v_{\text{th}} \mu_0^2\mu 
\frac{\left(1+\theta \right) ^{-5/2}}{\theta^{5/2}}
\left( 1+\alpha _{0}\right) \left[ -2\left( 90+231\theta +184\theta
^{2}+40\theta ^{3}\right) \right.  \nonumber \\
&&+3\mu\left( 1+\alpha _{0}\right) \left( 1+\theta \right) \left(
70+117\theta +44\theta ^{2}\right) -24\mu^{2}\left( 1+\alpha _{0}\right)
^{2}\left( 1+\theta \right) ^{2}\left( 5+4\theta \right)  \nonumber \\
&& \left. +30\mu^{3}\left( 1+\alpha _{0}\right) ^{3}\left(
1+\theta \right) ^{3}\right]  \;,  
\end{eqnarray}
\begin{eqnarray}
\label{a9}
\Omega_{02} &=&\frac{\sqrt{\pi}}{8}n\overline{\sigma}^2v_{\text{th}} \mu_0^2\mu 
\frac{\left(1+\theta \right) ^{-5/2}}{\theta^{1/2}}
\left( 1+\alpha_0\right) \left[ 2\left( 2+5\theta \right) +3\mu\left( 1+\alpha_0\right) \left( 1+\theta\right)\left(2+5\theta\right) \right.  \nonumber \\
& & \left. -24\mu^{2}\left( 1+\alpha_0\right) ^{2}\left(
1+\theta \right) ^{2}+30\mu^{3}\left( 1+\alpha _{0}\right) ^{3}\left(
1+\theta \right) ^{3}\right]  \;.  
\end{eqnarray}
The final expression of $c_0$ is obtained by substitution of Eq.\ (\ref{a1}) into Eq.\ (\ref{a6}) for the Gaussian thermostat and Eq.\ (\ref{a6.1}) for the stochastic thermostat and neglecting nonlinear terms in $c_0$ and $c$. The result is
\begin{equation}
\label{a10}
c_0=-\frac{\lambda_{00}+\lambda_{02}c+\frac{2}{15}\mu_0^{-2}\theta^2\left(\Omega_{00}+\Omega_{02}c\right)}
{\frac{1}{2}\lambda_{00}+\lambda_{01}+\frac{2}{15}
\mu_0^{-2}\theta^2\Omega_{01}},
\end{equation}
for the Gaussian thermostat while 
\begin{equation}
\label{a10.1}
c_0=-\frac{\lambda_{00}+\lambda_{02}c+\frac{2}{15}\mu_0^{-2}\theta^2\left(\Omega_{00}+\Omega_{02}c\right)}
{\lambda_{01}+\frac{2}{15}\mu_0^{-2}\theta^2\Omega_{01}},
\end{equation}
for the stochastic thermostat. 

Once the coefficient $c_0$ is given in terms of $\gamma=\mu_0/(\mu \theta)$ and the 
parameters of the mixture, the temperature ratio $\gamma$ can be explicitly 
obtained by numerically solving the condition (\ref{10.2.1}) for the Gaussian thermostat or the condition  (\ref{10.2.2}) for the stochastic thermostat.

\section{Second Sonine approximation}
\label{appB}

In this Appendix the coefficients $a_{1}$ and $b_{1}$ in the second Sonine approximation are determined. The knowledge of these coefficients allows one to get the diffusion $D$ and mobility $\lambda$ coefficients from Eqs.\ (\ref{20}), (\ref{21}), and (\ref{26}).  Let us consider first the case of the Gaussian thermostat (\ref{1}). For this thermostat, Eqs.\ (\ref{17}) and (\ref{18}) become, respectively 
\begin{equation}
\label{b1}
\frac{1}{2}\zeta\frac{\partial}{\partial {\bf v}}\cdot \left({\bf v}{\boldsymbol {\cal A}}\right)-J[{\boldsymbol {\cal A}},f]=-{\bf v} f_0^{(0)},
\end{equation}
\begin{equation}
\label{b2}
\frac{1}{2}\zeta\frac{\partial}{\partial {\bf v}}\cdot \left({\bf v}{\boldsymbol {\cal B}}\right)-J[{\boldsymbol {\cal B}},f]=
-\frac{1}{m_0}\frac{\partial}{\partial {\bf v}}f_0^{(0)}.
\end{equation}
Substitution of Eqs.\ (\ref{23}) and (\ref{23.1}) into the above integral equations gives  
\begin{equation}
\label{b3}
\frac{1}{2}\zeta\frac{\partial}{\partial {\bf v}}\cdot {\bf v}\left(
a_{1}f_{0,M}{\bf v}+a_{2}f_{0,M}{\bf S}_0\right)  -a_{1}J[f_{0,M}{\bf v},f]
 - a_{2}J[f_{0,M}{\bf S}_0 ,f]= -{\bf v} f_0^{(0)},
\end{equation}
\begin{equation}
\label{b4}
\frac{1}{2}\zeta\frac{\partial}{\partial {\bf v}}\cdot {\bf v}\left(
b_{1}f_{0,M}{\bf v}+b_{2}f_{0,M}{\bf S}_0 
\right)  -b_{1}J[f_{0,M}{\bf v},f]- b_{2}J[f_{0,M}{\bf S}_0,f]
 =-\frac{1}{m_0}\frac{\partial}{\partial {\bf v}}f_0^{(0)}.
\end{equation}
Next, Eqs.\ (\ref{b3}) and (\ref{b4}) are multiplied by ${\bf v}$ and 
integrated over the velocity. The result is 
\begin{equation}
\label{b5}
(\nu_1-\frac{1}{2}\zeta)a_{1}+\nu_2 a_{2}=-1,
\end{equation}
\begin{equation}
\label{b6}
(\nu_1-\frac{1}{2}\zeta)b_{1}+\nu_2 b_{2}=T_0^{-1},
\end{equation}
where I have introduced the quantities
\begin{equation}
\label{b7}
\nu_1=-\frac{m_0}{3n_0T_0}\int\, d{\bf v}\, {\bf v}\cdot J[f_{0,M}{\bf v},f],
\end{equation}
\begin{equation}
\label{b8}
\nu_2=-\frac{m_0}{3n_0T_0}\int\, d{\bf v}\, {\bf v}\cdot J[f_{0,M}{\bf S}_0 ,f].
\end{equation}
If only the first Sonine correction is retained (which means $a_{2}=b_2=0$), the solution to Eqs.\ (\ref{b5}) and (\ref{b6}) yields $a_1=-T_0b_1$, and so  the modified Einstein relation  (\ref{n1}) applies ($\epsilon\to 1$).

To close the problem in the second Sonine approximation, one multiplies Eqs.\ (\ref{b3}) and (\ref{b4}) by 
${\bf S}_0({\bf v})$ and integrates over the velocity. Following identical mathematical steps as those made before, one gets
\begin{equation}
\label{b9}
\left(\nu_3-\zeta T_0^{-1}\right)a_{1}+\left(\nu_4-\frac{3}{2}\zeta\right) a_{2}=-\frac{1}{2}c_0T_0^{-1},
\end{equation}
\begin{equation}
\label{b10}
\left(\nu_3-\zeta T_0^{-1}\right)b_{1}+\left(\nu_4-\frac{3}{2}\zeta\right) b_{2}=0,
\end{equation}
where I have introduced the quantities
\begin{equation}
\label{b11}
\nu_3=-\frac{2}{15}\frac{m_0}{n_0T_0^3}\int\, d{\bf v}\, {\bf S}_0 
\cdot J[f_{0,M}{\bf v},f],
\end{equation}
\begin{equation}
\label{b12}
\nu_4=-\frac{2}{15}\frac{m_0}{n_0T_0^3}\int\, d{\bf v}\, {\bf S}_0 
\cdot J[f_{0,M}{\bf S}_0,f_].
\end{equation}

In reduced units and by using matrix notation, Eqs.\  (\ref{b5}) and (\ref{b9}), along with Eqs.\  (\ref{b6}) and (\ref{b10}), can be rewritten, respectively,  as 
\begin{equation}
\label{b13}
\left(
\begin{array}{cc}
\nu_1^*-\case{1}{2}\zeta^{*}& \nu_2^*\\
\nu_3^*-\zeta^{*}& \nu_4^*-\case{3}{2}\zeta^{*}
\end{array}
\right)
\left(
\begin{array}{c}
a_{1}^*\\
a_{2}^*
\end{array}
\right)
=-
\left(
\begin{array}{c}
1\\
c_0/2
\end{array}
\right),
\end{equation}
\begin{equation}
\label{b14}
\left(
\begin{array}{cc}
\nu_1^*-\case{1}{2}\zeta^{*}& \nu_2^*\\
\nu_3^*-\zeta^{*}& \nu_4^*-\case{3}{2}\zeta^{*}
\end{array}
\right)
\left(
\begin{array}{c}
b_{1}^*\\
b_{2}^*
\end{array}
\right)
=
\left(
\begin{array}{c}
1\\
0
\end{array}
\right).
\end{equation}
Here, $\zeta^*=\zeta/\nu$, $\nu_1^*=\nu_1/\nu$, $\nu_2^*=\nu_2/T_0\nu$, $\nu_3^*=T_0\nu_3/\nu$, and $\nu_4^*=\nu_4/\nu$ with $\nu=n\sigma^2v_{\text{th}}$. Further, $a_{1}^*=\nu a_{1}$, $a_{2}^*=T_0\nu a_{2}$, $b_1^*=T_0\nu b_1$, and $b_2^*=T_0^2\nu b_2$. Equation (\ref{b13}) coincides with the one recently derived in the free cooling case \cite{GM04}. The solution to Eqs.\ (\ref{b13}) and (\ref{b14}) provides the explicit expression of the Einstein ratio in the second Sonine approximation. The result is 
\begin{equation}
\label{b15}
\epsilon=\frac{D}{T_0\lambda}=-\frac{a_1^*}{b_1^*}=1-\frac{c_0}{2}\frac{\nu_2^*}{\nu_4^*-\frac{3}{2}\zeta^*}.
\end{equation}

The analysis in the case of the stochastic thermostat (\ref{3}) is similar to that made above for the Gaussian one. The linear integral equations are now given by 
\begin{equation}
\label{b16}
-\frac{1}{2}\frac{T}{m}\zeta \left(\frac{\partial}{\partial {\bf v}}\right)^2
{\boldsymbol {\cal A}}-J[{\boldsymbol {\cal A}},f]=-{\bf v} f_0^{(0)},
\end{equation}
\begin{equation}
\label{b17}
-\frac{1}{2}\frac{T}{m}\zeta \left(\frac{\partial}{\partial {\bf v}}\right)^2
{\boldsymbol {\cal B}}-J[{\boldsymbol {\cal B}},f]=
-\frac{1}{m_0}\frac{\partial}{\partial {\bf v}}f_0^{(0)}.
\end{equation}
The corresponding matrix equations defining the Sonine coefficients $\{a_1, a_2, b_1, b_2\}$ can be obtained by multiplying Eqs.\ (\ref{b16}) and (\ref{b17}) by ${\bf v}$ and ${\bf S}_0({\bf v})$. After some algebra, one gets
\begin{equation}
\label{b18}
\left(
\begin{array}{cc}
\nu_1^*& \nu_2^*\\
\nu_3^*-\zeta_0^{*}& \nu_4^*
\end{array}
\right)
\left(
\begin{array}{c}
a_{1}^*\\
a_{2}^*
\end{array}
\right)
=-
\left(
\begin{array}{c}
1\\
c_0/2
\end{array}
\right),
\end{equation}
\begin{equation}
\label{b19}
\left(
\begin{array}{cc}
\nu_1^*& \nu_2^*\\
\nu_3^*-\zeta_0^{*}& \nu_4^*
\end{array}
\right)
\left(
\begin{array}{c}
b_{1}^*\\
b_{2}^*
\end{array}
\right)
=
\left(
\begin{array}{c}
1\\
0
\end{array}
\right),
\end{equation}
where $\zeta_0^*=\zeta_0/\nu$ and use has been made of the relation 
\begin{equation}
\label{b19bis}
\frac{\zeta T}{m}=\frac{\zeta_0T_0}{m_0}.
\end{equation}
The solution to Eqs.\ (\ref{b18}) and (\ref{b19}) leads to the following expression for the Einstein ratio: 
\begin{equation}
\label{b20}
\epsilon=1-\frac{c_0}{2}\frac{\nu_2^*}{\nu_4^*}.
\end{equation}

The collision integrals $\nu_2^*$ and $\nu_4^*$ appearing in the Einstein ratios (\ref{b15}) and (\ref{b20}) have been recently evaluated by the author \cite{GM04}. They are given by  
\begin{equation}
\label{b21}
\nu_2^*=\frac{2}{3}\sqrt{\pi}\mu\left(\frac{\overline{\sigma}}
{\sigma}\right)^2(1+\alpha_0)\left[\theta(1+\theta)\right]^{-1/2}
\left[1+\frac{3}{16}c\left(\frac{\theta}{1+\theta}\right)^{2}\right],
\end{equation}
\begin{equation}
\label{b22}
\nu_4^*=\frac{2}{15}\sqrt{\pi}\mu
\left(\frac{\overline{\sigma}}{\sigma}\right)^2(1+\alpha_0)
\left(\frac{\theta}{1+\theta}\right)^{3/2} 
\left(A-5\frac{1+\theta}{\theta}B\right), 
\end{equation} 
where  
\begin{eqnarray}
\label{b23}
A&=& 2\mu^2\left(\frac{1+\theta}{\theta}\right)^2
\left(2\alpha_0^{2}-3\alpha_0+4\right)
(8+5\theta)\nonumber\\
& & 
-\mu(1+\theta)\left[2\beta\theta^{-2}(8+5\theta)(7\alpha_0
-11)+2\theta^{-1}(29\alpha_0-37)-25(1-\alpha_0)\right]
\nonumber\\
& & +18\beta^2\theta^{-2}(8+5\theta)+
2\beta\theta^{-1}(25+66\theta)+5\theta^{-1}
(6+11\theta)-5(1+\theta)\theta^{-2}(6+5\theta)
\nonumber\\
& & +\frac{c}{16}\left(1+\theta\right)^{-2}\left\{15\theta^3
\mu(1+\alpha_0)(4\mu(1+\alpha_0)-5)
\right.\nonumber\\
& & +2\left[45+540\mu_0^2+16
\mu(\alpha_0-36)
+4\mu^2(134+5\alpha_0+6\alpha_0^2)\right.
\nonumber\\
& & \left.
-4\mu_0(148+\mu(7\alpha_0-263))\right]+
\theta^2\left[-30-\mu(267+217\alpha_0)
\right.\nonumber\\
& & \left.
+14\mu^2
(17+29\alpha_0+12\alpha_0^2)
+10\mu_0(7\mu(1+\alpha_0)-5)\right]
\nonumber\\
& & 
+\theta\left[-315+270\mu_0^2-2\mu(55\alpha_0+57)
+\mu^2(440+326\alpha_0+156\alpha_0^2)\right.\nonumber\\
& & \left.\left.
+2\mu_0(-2+\mu(7\alpha_0+277))\right]\right\},
\end{eqnarray}
\begin{eqnarray}
\label{b24}
B&=&5(1+2\beta)+\mu(1+\theta)\left[5(1-\alpha_0)-2(7\alpha_0-11)
\beta \theta^{-1}\right]+18\beta^2\theta^{-1} \nonumber\\
& & +2\mu^2(2\alpha_0^2-3\alpha_0+4)\theta^{-1}(1+\theta)^2-
5\theta^{-1}(1+\theta)\nonumber\\
& & +\frac{c}{16}\frac{\theta}{(1+\theta)^2}\left\{3\theta^2\mu(1+
\alpha_0)\left[4\mu(1+\alpha_0)-5\right]+\theta\left[
2\mu_0\left(7\mu(1+\alpha_0)-5\right)\right.\right.
\nonumber\\
& & \left. 
+\mu\left(-5(9+7\alpha_0)+\mu(38+62\alpha_0+24\alpha_0^2)
\right)\right]-15+54\mu_0^2-20\mu(3+\alpha_0)\nonumber\\
& & \left. +2\mu^2(40+19\alpha_0+6\alpha_0^2)+2\mu_0\left[
\mu(61+7\alpha_0-20)\right]\right\}.
\end{eqnarray}
In the above expressions, $\beta=\mu_0-\mu\theta=\mu_0\left(1-\gamma^{-1}\right)$.

\end{document}